\documentclass[aps, prl, reprint, superscriptaddress, showpacs]{revtex4-1}
 \usepackage{graphicx}
 \usepackage{amsmath}
 \usepackage[colorlinks=true, allcolors=blue]{hyperref}
\usepackage{graphicx} 

\begin{document}
\title{Three-dimensional imaging of oxygen dopant distribution in Sr$_2$CuO$_{3+\delta}$ by electron ptychography}

\author{Hongbin Yang}
  \affiliation{School of Applied and Engineering Physics, Cornell University, Ithaca, NY 14853, USA}

\author{Jinkwon Kim}
  \affiliation{Department of Materials Science and Engineering, Cornell University, Ithaca, NY 14853, USA}

\author{Desheng Ma}
 \affiliation{School of Applied and Engineering Physics, Cornell University, Ithaca, NY 14853, USA}

\author{Dasol Yoon}
 \affiliation{Department of Materials Science and Engineering, Cornell University, Ithaca, NY 14853, USA}

\author{Darrell G. Schlom}
  \affiliation{Department of Materials Science and Engineering, Cornell University, Ithaca, NY 14853, USA}
  \affiliation{Kavli Institute at Cornell for Nanoscale Science, Cornell University, Ithaca, NY 14853, USA}
  \affiliation{Leibniz-Institut f{\"u}r Kristallz{\"u}chtung, Max-Born-Stra{\ss}e 2, 12489 Berlin, Germany}

\author{David A. Muller}
\email{david.a.muller@cornell.edu}
  \affiliation{School of Applied and Engineering Physics, Cornell University, Ithaca, NY 14853, USA}
  \affiliation{Kavli Institute at Cornell for Nanoscale Science, Cornell University, Ithaca, NY 14853, USA}

\begin{abstract}
Oxygen dopants play a critical role in tuning the properties of cuprate superconductors, yet it is challenging to visualize them at the atomic scale. 
Here, we use multislice electron ptychography to directly image oxygen dopants in a Sr$_2$CuO$_{3+\delta}$ film. We observe oxygen dopants at interstitial sites between the Cu–O chains, with a strong preference for clustering in tensile-strained regions, which are often associated with dislocations and interfacial steps. These findings indicate that the oxygen dopant distribution in cuprates is not random but rather sensitive to strain field, suggesting strain as a doping tuning parameter.
\end{abstract}

\maketitle
Electronic and magnetic properties of materials are heavily influenced by doping. Among the most versatile and ubiquitous dopants is oxygen. In complex oxides, especially high-$T_{c}$ superconductors, oxygen atoms not only form the structural framework but also govern the charge concentration and emergent electronic phases. While the effect of doping can be inferred from transport and spectroscopy measurements, direct imaging of oxygen dopants has been challenging. Imaging the distribution and local atomic structure of dopant atoms will help understand electronic inhomogeneity in cuprates \cite{pan_microscopic_2001, Johnson_YBCO_2008} and the effect of disorder on superconductivity \cite{fujita_effect_2005}. Sr$_2$CuO$_{3+\delta}$ is a cuprate with one-dimensional Cu-O chains \cite{Teske_1969, Weller_1989}, offering an ideal model system for investigating the oxygen interstitials. Oxygen doping of Sr$_2$CuO$_{3+\delta}$ results in exotic electronic and magnetic excitations \cite{schlappa_spinorbital_2012, chen_anomalously_2021}, and under extreme conditions, high-$T_{c}$ superconductivity \cite{hiroi_new_1993, liu_enhancement_2006, geballe_enhanced_2009}.

Imaging of dopant atoms is feasible with annular dark-field (ADF) imaging in a scanning transmission electron microscope (STEM), provided the atomic number difference is large. For example, ADF imaging has observed Sb dopants in Si \cite{voyles_atomic-scale_2002}, Hf atoms in SiO$_2$ \cite{van_benthem_three-dimensional_2005}, A-site dopants and vacancies in SrTiO$_3$ \cite{hwang_three-dimensional_2013, kim_direct_2016}, Ce dopants in BN\cite{ishikawa_three-dimensional_2020}, and many others \cite{saito_three-dimensional_2017, johnson_unusual_2019}. Using phase sensitive imaging techniques such as differential phase contrast \cite{lazic_phase_2016}, oxygen atom columns can be observed, for example in Bi$_2$Sr$_2$CaCu$_2$O$_{8+x}$ \cite{song_visualization_2019} and La$_2$CuO$_4$ \cite{zhang_visualization_2022}. Unfortunately, results from these techniques are often complicated by multiple scattering and channeling of fast electrons, especially in samples thicker than a few nanometers \cite{voyles_depth-dependent_2004, xin_depth_2008, karapetyan_3d_2026}, limiting their use in imaging dopants in crystalline host materials. With the advent of multislice electron ptychography (MEP) \cite{jiang_electron_2018, chen_electron_2021}, both heavy and light atoms can be imaged with significantly improved lateral and depth resolution, and crucially without the limits from multiple scattering. Recently, MEP has been applied to study various point defects, including dopants of heavy elements \cite{chen_imaging_nodate, dong_sub-nanometer_2025}, oxygen vacancies and interstitials in nickelates \cite{dong_visualization_2024, dong_interstitial_2025}, oxygen interstitials in an alloy \cite{liu_direct_2023}, as well as vacancies in SiC \cite{kim_quantifying_2025} and lithium vacancies \cite{yoon_atomic-scale_2026}. The three-dimensional (3D) imaging capability and light atom sensitivity of MEP offer new opportunities for imaging and even quantify oxygen dopants in cuprates.

Here, we demonstrate quantitative imaging of oxygen dopants in Sr$_2$CuO$_{3+\delta}$ with MEP. In a Sr$_2$CuO$_{3+\delta}$ film on SrTiO$_3$ substrate, we observe oxygen interstitial dopants located between the Cu-O chains. We develop a method for quantification of oxygen dopants by referencing them with apical oxygen of the host lattice. This is applied to count the number of interstitials observed at each site. We further image the strain, dislocations, and interface roughness in 3D and evaluate their influence on the distribution of oxygen interstitials. Strain analysis reveals a local expansion of Cu-to-Cu spacing at the location of oxygen interstitials, whose magnitude is proportional to the amount of oxygen interstitials.


\begin{figure}[t]
\includegraphics[width=8.6cm]{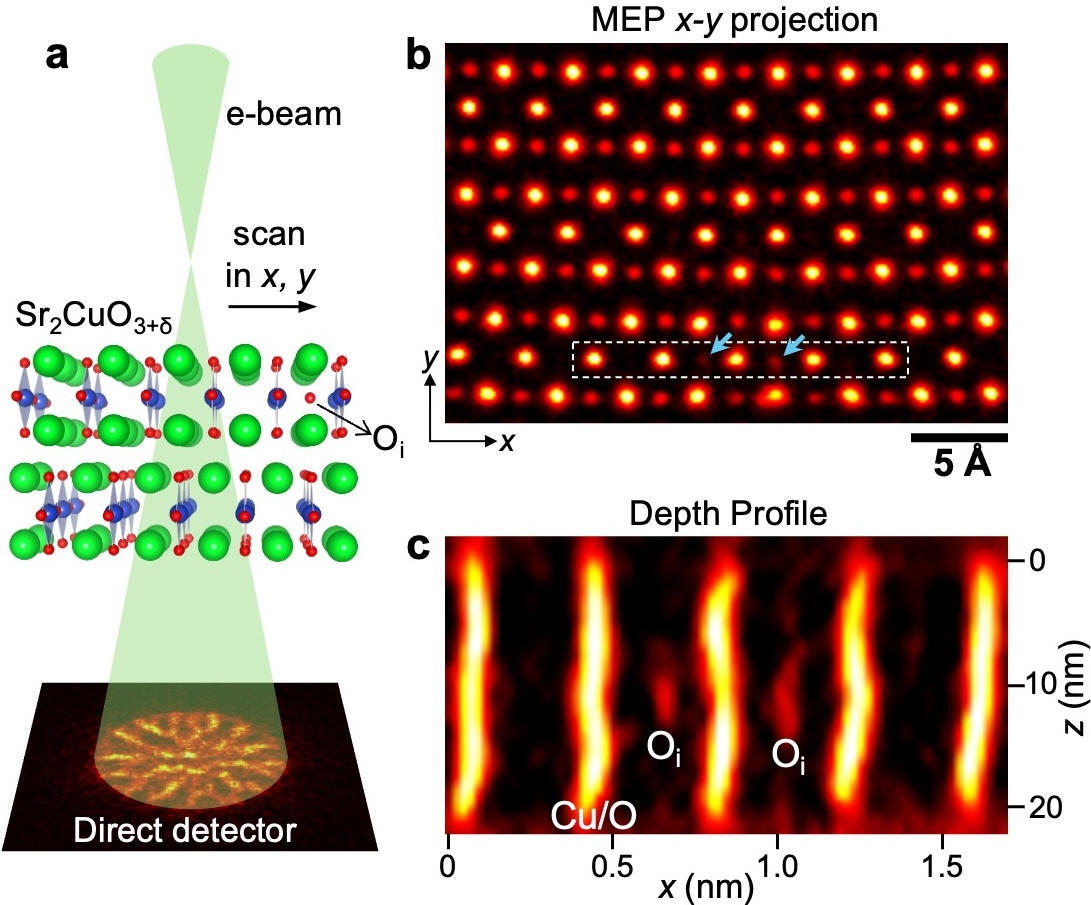}
\caption{Oxygen interstitials in Sr$_2$CuO$_{3+\delta}$ observed by MEP. (a) Schematic of a scanning electron diffraction experiment. Green, blue, and red spheres represent Sr, Cu, and O atoms, respectively. (b) A MEP image of Sr$_2$CuO$_{3+\delta}$ along the [010] zone axis. Brighter/yellow atom columns are Sr and Cu/O; dimmer/red columns are apical oxygens (O$\mathrm{_{ap}}$). Interstitial oxygens (O$\mathrm{_i}$) are also visible at some sites, as indicated by the arrows. (c) Depth profile across along the five Cu/O columns marked in (b). The two prominent O$\mathrm{_i}$ between the Cu/O are labeled. Lateral displacements of Cu/O columns are also visible near O$\mathrm{_i}$.}
\label{fig:MEP}
\end{figure}

We grow Sr$_2$CuO$_{3+\delta}$ films on a SrTiO$_3$ substrate by molecular beam epitaxy \cite{SM}. The film was capped by a few nanometers of Al$_2$O$_3$ for better stability in ambient condition. Cross-section TEM lamellas of the heterostructure were prepared by focused ion beam. We perform STEM imaging of Sr$_2$CuO$_{3+\delta}$ at 300 kV and acquire 4D-STEM data for MEP reconstructions. As illustrated in Fig. \hyperref[fig:MEP]{\ref{fig:MEP}(a)}, an overfocused electron beam is scanned across the sample while diffraction patterns are recorded with a pixelated direct detector (EMPAD G2) \cite{philipp_very-high_2022}. We then use a multislice algorithm in the \texttt{fold\_slice} package \cite{chen_electron_2021} to iteratively solve the object and probe wavefunctions. The reconstructed object phase image in Fig. \hyperref[fig:MEP]{\ref{fig:MEP}(b)} shows the atomic structure of Sr$_2$CuO$_{3+\delta}$ along the [010] zone axis. 

Sr$_2$CuO$_{3+\delta}$ is a layered cuprate with one-dimensional corner-sharing Cu-O chains \cite{Teske_1969, Weller_1989}. Imaging along the [010] direction allows the chains to be viewed end-on. The Sr and alternating Cu and O (Cu/O) columns appear with brighter contrast due to their larger atomic number, while the apical oxygen columns appear dimmer. The vacant equatorial oxygen lattice sites between neighboring Cu/O columns along the x-direction appear as background intensity in the MEP x-y projection image.

While most equatorial oxygen lattice sites between the Cu-O chains are vacant, we observe interstitial oxygens (O$\mathrm{_i}$) at several locations, with the most prominent ones indicated by the arrows in Fig. \hyperref[fig:MEP]{\ref{fig:MEP}(b)}. The O$\mathrm{_i}$ are visible also from the MEP depth profile in Fig. \hyperref[fig:MEP]{\ref{fig:MEP}(c)}, which also reveal their positions along the depth ($z$) direction. Notably, the presence of O$\mathrm{_i}$ is accompanied by lateral displacements of the neighboring Cu/O columns. In addition to the two labelled O$\mathrm{_i}$ columns in Fig. \hyperref[fig:MEP]{\ref{fig:MEP}(b, c)}, there are several other interstitial sites that exhibit weaker intensity, which we attribute to fewer O$\mathrm{_i}$ atoms at these positions.

\begin{figure}[t]
\includegraphics[width=8.6cm]{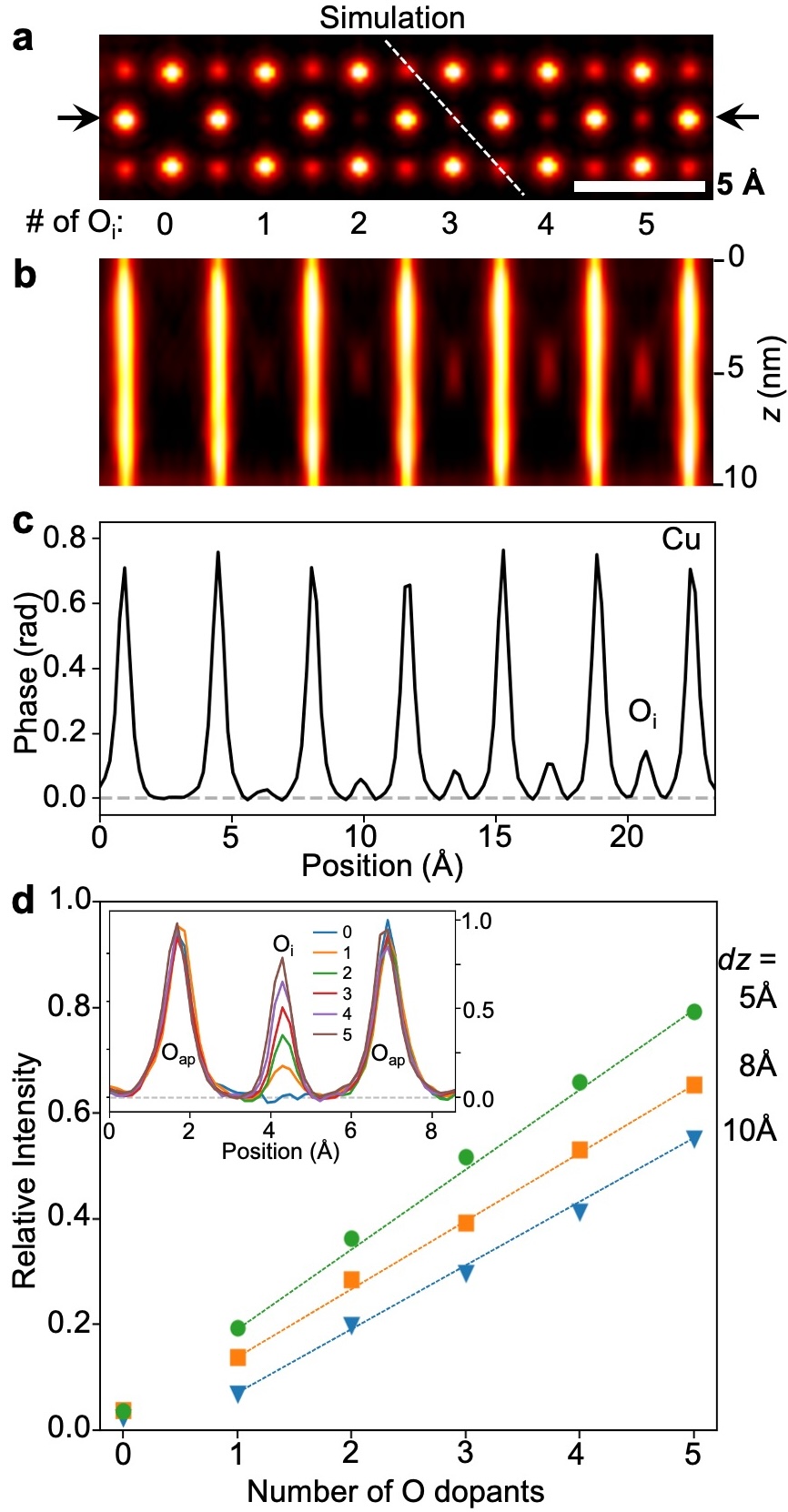}
\caption{Simulations for quantification. (a) Simulated MEP image of an artificial Sr$_2$CuO$_{3+\delta}$ lattice with 0 to 5 interstitial oxygens per atomic column. (b) depth profile of the Cu-O$\mathrm{_i}$ atomic plane. (c) Line profile across the Cu-O$\mathrm{_i}$ plane in (a). (d) Relative intensity as functions of interstitial oxygen content and slice thickness $dz$ in MEP reconstructions. The inset shows the intensity profile of apical-interstitial-apical oxygen along the dashed line in (a).}
\label{fig:simulation}
\end{figure}

To quantify the number of interstitial oxygen atoms from the MEP image contrast, we performed MEP reconstruction on simulated 4D-STEM data. We built a structure model that contains a varying number of oxygens at the interstitial sites along the depth direction (Figure S1). The reconstructed MEP images in Fig. \hyperref[fig:simulation]{\ref{fig:simulation}(a, b)} show that the intensity of O$\mathrm{_i}$ scales with the number of interstitial oxygen atoms along $z$. The image intensity produced by five O$\mathrm{_i}$ along $z$ appears as strong as that of the apical oxygens, while a single O$\mathrm{_i}$ is barely visible but can be resolved from the line profile in  Fig. \hyperref[fig:simulation]{\ref{fig:simulation}(c)}, and show clear difference with the zero O$\mathrm{_i}$ vacancy site.

\begin{figure*}
\includegraphics[width=17.2cm]{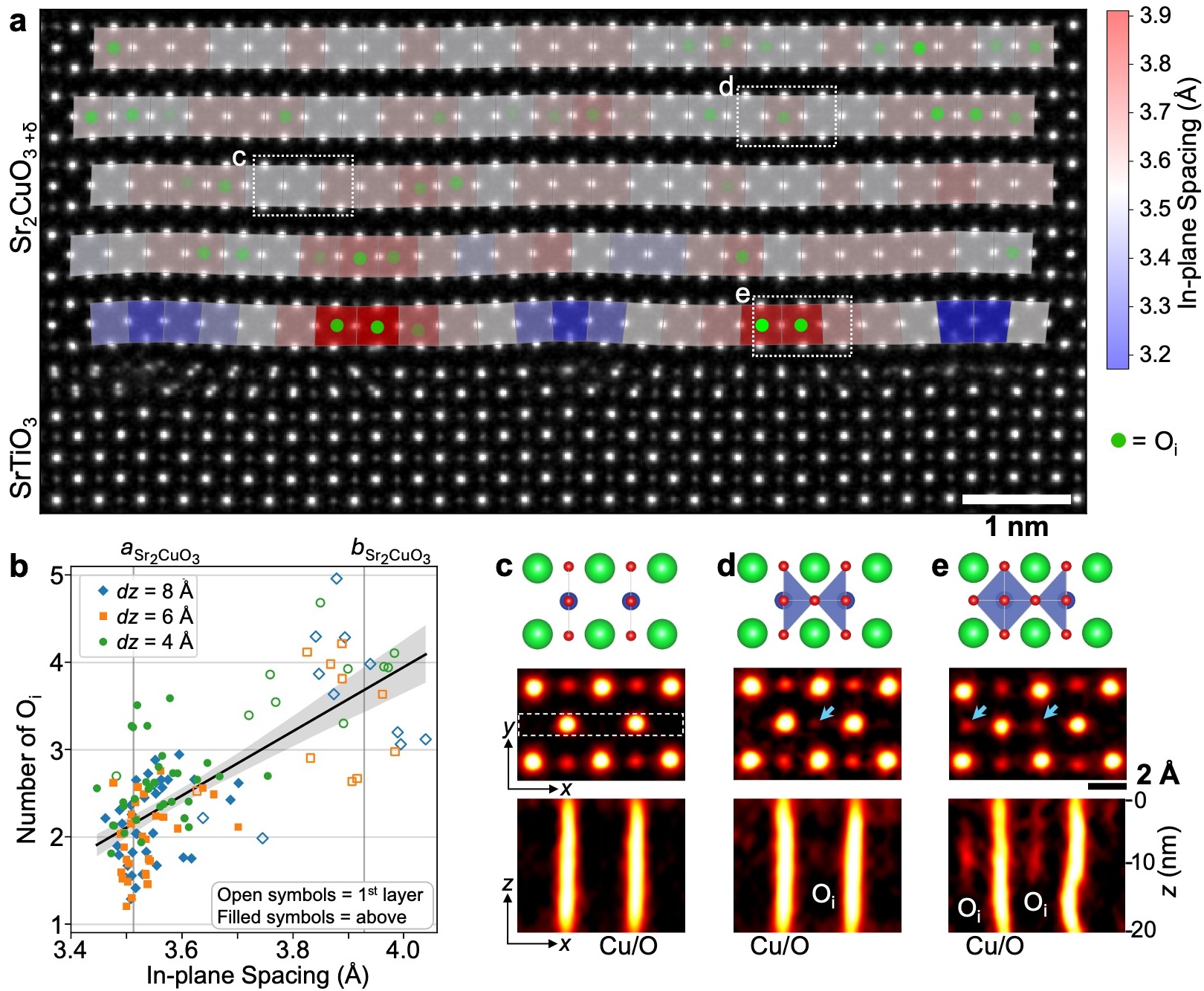}
\caption{Doping-strain correlation. (a) A large field-of-view MEP image of Sr$_2$CuO$_{3+\delta}$/SrTiO$_3$. The in-plane lattice spacing (colored blocks) and oxygen interstitial locations (green dots) are also shown. The transparency of the green dots represent the local O$\mathrm{_i}$ intensity. (b) The number of interstitial oxygen along $z$ as a function of in-plane Cu-Cu spacing obtained from $dz$ = 8, 6, 4 \AA{} reconstructions. Open symbols correspond to O$\mathrm{_i}$ in the first layers of Sr$_2$CuO$_{3+\delta}$, filled symbols correspond to O$\mathrm{_i}$ in other layers. (c, d, e) Structural model (top), MEP image (middle), and depth profile (bottom) for Sr$_2$CuO$_{3+\delta}$ with varying numbers of O$\mathrm{_i}$. The depth profiles of Cu-O planes are obtained from the dashed rectangle in the corresponding middle panels of (c, d, e). The oxygen interstitials are indicated by the arrows.}
\label{fig:correlation}
\end{figure*}

The relatively weak intensity of O$\mathrm{_i}$ columns in the MEP images is due in part to the limited depth resolution of 2 to 3 nanometers. The intensity of atomic columns in each slice of a MEP reconstruction reflects the projected electrostatic potential integrated over the slice thickness $dz$. For periodically spaced host-lattice atoms, the projected potential scales linearly with the number of atoms within $dz$. In contrast, dopant atoms are localized at specific depth; if the depth resolution is coarser than the actual depth extent of the O$\mathrm{_i}$ atoms, the signal is diluted over multiple slices, reducing the peak intensity. Therefore, the projected potentials of O$\mathrm{_i}$ have a different dependence on $dz$ than those of the host lattice.

This difference in $dz$ dependence means that the visibility of dopant atoms is sensitive to $dz$ in MEP reconstructions. In Fig. \hyperref[fig:simulation]{\ref{fig:simulation}(d)} we plot the intensity of interstitial relative to the apical lattice oxygens. This relative intensity is defined by I$_{O_i}$/I$_{O_{ap}}$, where I$_{O_i}$ and I$_{O_{ap}}$ are phase intensity of interstitial and apical oxygen atoms, respectively. They are obtained from the diagonal line profiles (path indicated by the dashed line in Fig. \hyperref[fig:simulation]{\ref{fig:simulation}(a)}) shown in the inset of Fig. \hyperref[fig:simulation]{\ref{fig:simulation}(d)}. It is clear that the smaller $dz$ offers higher intensity of O$\mathrm{_i}$ relative to O$\mathrm{_{ap}}$. The relative intensity thus offers a more reliable means to quantify the number of O$\mathrm{_i}$ than the absolute reconstructed object phase.

Next, we come back to our experimental data to study the O$\mathrm{_i}$ spatial distribution in the the Sr$_2$CuO$_{3+\delta}$ film. We identify O$\mathrm{_i}$ with three criteria \cite{voyles_imaging_2003}. Firstly, the phase intensity of the O$\mathrm{_i}$ candidates should exceed the noise level. Due to the weak phase of O$\mathrm{_i}$, we find that it is necessary to subtract a slice-dependent background in MEP reconstructions \cite{bhat_sensitivity_2026}, which can be obtained from the regions between atom columns (in this case, the Ruddlesden-Popper gap between the layers). Secondly, the O$\mathrm{_i}$ atomic column should have a minimum size in both the lateral and depth directions. And thirdly, the lateral locations of O$\mathrm{_i}$ should be near the midpoint between two neighboring Cu/O columns. We also exclude the dopant-like atoms near the top and bottom surfaces of the lamella, which are prone to surface reconstructions or absorbates introduced during to TEM sample preparation.

Applying these criteria, we located the O$\mathrm{_i}$ in the Sr$_2$CuO$_{3+\delta}$ film across a large field of view. The location of O$\mathrm{_i}$, their relative intensity, and the local lattice spacing are overlaid on the MEP image in Fig. \hyperref[fig:correlation]{\ref{fig:correlation}(a)}. The more prominent O$\mathrm{_i}$ are found near the interface, in the first perovskite layer of Sr$_2$CuO$_{3+\delta}$, which exhibits alternating compressive and tensile strain along the $b$-axis ($x$ direction, perpendicular to the Cu-O chain). Away from the interface, the strain relaxes quickly; the amount of O$\mathrm{_i}$ at each site also reduces.

\begin{figure}
\includegraphics[width=8.6cm]{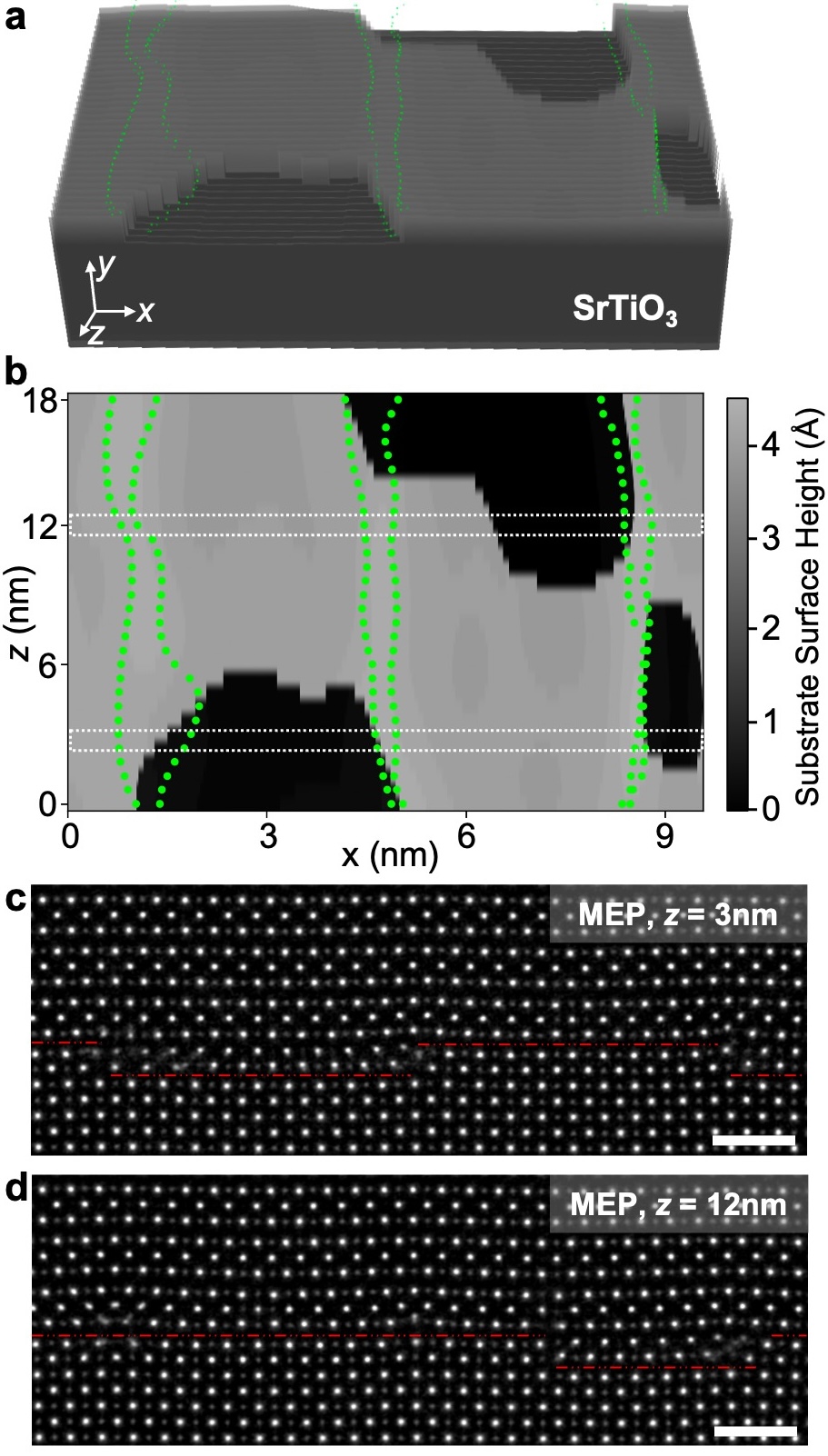}
\caption{Dislocation and interface roughness in 3D. (a) A 3D rendering of the STO substrate after film growth, showing 4 \AA{} height interface steps and dislocations along the step edges. (b) Top-down (substrate surface normal direction/$y$-axis in MEP) view of interface roughness and dislocations. The dislocation cores are indicated with green dots. (c, d) MEP images at $z$ = 3 and 12 nm, as marked in (b). The vdW gap between Sr$_2$CuO$_{3+\delta}$ and SrTiO$_3$ is marked with red dashed lines. Scalebars: 1 nm.}
\label{fig:3d}
\end{figure}

Figure \hyperref[fig:correlation]{\ref{fig:correlation}(b)} shows a positive correlation between the O$\mathrm{_i}$ count per site and the local Cu-Cu spacing. The number of O$\mathrm{_i}$ at each site is estimated according to the calibration in Fig. \hyperref[fig:simulation]{\ref{fig:simulation}(d)}. Reconstructing the same 4D-STEM dataset with $dz$ = 4, 6, 8 \AA{} yields similar correlations between doping and strain. The distribution reveals two preferred lattice spacings. Unit cells under moderate tensile strain (spacing below 3.6 \AA{}) tend to host a small number of O$\mathrm{_i}$ in each column, typically between 1 and 3. In contrast, the sites that accommodate O$\mathrm{_i}$ clusters have a spacing close to 3.9 \AA{}. These values are close to the lattice parameter of orthorhombic Sr$_2$CuO$_{3+\delta}$ ($a$ = 3.928 \AA{}, $b$ = 3.512 \AA{}), which has a large lattice mismatch with cubic SrTiO$_3$ ($a$ = $b$ = 3.905 \AA{}) in the $b$-axis.

Figures \hyperref[fig:correlation]{\ref{fig:correlation}(c, d, e)} present the zoomed-in images and depth profiles for the three representative sites that contain 0, 3, and nearly 5 O$\mathrm{_i}$ within the depth of field (2 to 3 nm), respectively. The presence of multiple O$\mathrm{_i}$ clusters results in expanded spacing between the surrounding Cu/O columns, as seen clearly in Fig. \hyperref[fig:correlation]{\ref{fig:correlation}(e)}, a feature that is similar to our earlier observation in Fig. 1 (c). 


In addition to oxygen interstitials and strain, MEP has resolved interface roughness and edge dislocations in 3D. As a layered oxide, the Sr$_2$CuO$_{3+\delta}$ film forms a van der Waals (vdW) gap with the SrTiO$_3$ substrate, as visible from the MEP images of the Sr$_2$CuO$_{3+\delta}$/SrTiO$_3$ interface in Fig. 3(a). Tracking along the vdW gap, we observe the substrate surface exhibit a roughness of about 4 \AA{} (one SrTiO$_3$ unit cell) in height. From inverse fast Fourier transform (iFFT)  of the (001) reflections of SrTiO$_3$ substrate, we extract the vdW gap position as a function of depth (Figure S2). Figure \hyperref[fig:3d]{\ref{fig:3d}(a)} shows the resulting 3D surface model of the SrTiO$_3$ substrate after film growth, revealing several surface steps with 4\AA{} height. A top-down view of the surface topology in Fig. \hyperref[fig:3d]{\ref{fig:3d}(b)} reveals additional picometer-level roughness arising from small variations in the local bond distortion.

The lattice mismatch also gives arise to misfit dislocations. We find the dislocation core by iFFT of the (200) reflection and then locate the singularities in the geometric phase map (Figure S2). The extra atomic planes associated with the dislocation are terminated by the dislocation cores near the interface, as indicated by green dots in Fig. \hyperref[fig:3d]{\ref{fig:3d}(a, b)}. The top-down view in Fig. \hyperref[fig:3d]{\ref{fig:3d}(b)} shows more clearly that the dislocation cores appear in pairs and tend to locate near the step edges of the SrTiO$_3$ surface. The location of dislocation cores also coincide with the compressive strained regions in Fig. \hyperref[fig:correlation]{\ref{fig:correlation}(a)}, as a result of the wrinkling of the first layer of the Sr$_2$CuO$_{3+\delta}$ film. At the troughs of the SrTiO$_3$ surface, we observe a local Sr$_3$Cu$_2$O$_{5+\delta}$ intergrowth, a $n$ = 2 member of the general Sr$_{n+1}$Cu$_n$O$_{2n+1+\delta}$ series \cite{hiroi_new_1993}, as shown in Fig. \hyperref[fig:3d]{\ref{fig:3d}(c, d)}. A similar effect has been observed for Sr$_2$RuO$_4$ films \cite{kim_superconducting_2021}. This secondary phase effectively eliminates the substrate surface roughness for subsequent growth of the Sr$_2$CuO$_{3+\delta}$ film. Taken together, the oxygen dopant distribution in Sr$_2$CuO$_{3+\delta}$ is influenced by not only strain, but also by interface roughness and dislocations.

When imaging beam-sensitive materials such as cuprates, radiation damage must be considered. In the present study, we used a moderate dose of $1.2\times 10^{5}$ electrons/\AA{}$^2$ to avoid significant modifications to the Sr$_2$CuO$_{3+\delta}$ film. Under this condition, our MEP experimental results reveal that the minimum detectable O$\mathrm{_i}$ signal corresponds to two interstitials per column. This observation is consistent with our dose-dependent simulations (Figure S3), which indicate that a total dose of $1\times 10^{7}$ electrons/\AA{}$^2$ would be ideal to clearly identify a single O$\mathrm{_i}$ from the background noise. Thus, while our imaging conditions are sufficient to identify small interstitial clusters, we do not rule out the presence of isolated single O$\mathrm{_i}$ atoms below our detection limit.

We next discuss the origin of the oxygen interstitials. The O$\mathrm{_i}$ observed here were not deliberately introduced through post-growth annealing in oxygen or ozone. Instead, we attribute their presence primarily to the lattice-mismatch-driven strain relaxation. By incorporating extra oxygen at the interstitial sites, the Sr$_2$CuO$_{3+\delta}$ film accommodates the tensile strain from SrTiO$_3$ substrate. This interpretation is supported by our observation that O$\mathrm{_i}$ are preferentially located at the tensile strained regions, and are most abundant near the interface, where the mismatch is largest. Away from the interface, MEP results suggest that $\delta$ $\approx$ 0.005 for this Sr$_2$CuO$_{3+\delta}$ film, a relatively small amount compared to the intentionally annealed films in literature \cite{chen_anomalously_2021, karimoto_new_2001}.

An alternative scenario is that some regions with O$\mathrm{_i}$ are accompanied by nearby oxygen vacancies, effectively forming sub-nanometer scale orthorhombic twin domains. In this situation, the stoichiometry and electronic structure may not be significantly altered, but the length and direction of the Cu-O chains would be different from the expectation from domain size. Further improvement in depth resolution of MEP would be helpful for definitively distinguishing between the two scenarios.

In summary, we have demonstrated three-dimensional imaging and quantification of oxygen dopants in Sr$_2$CuO$_{3+\delta}$ using multislice electron ptychography. The oxygen dopants preferentially cluster in tensile-strained regions, with their amount correlating with local lattice spacing. MEP imaging further resolved misfit dislocations and interface roughness, revealing a rich landscape of structural heterogeneity. This work establishes MEP as a powerful tool for studying light-element dopants in complex oxides and opens new avenues for understanding doping inhomogeneity in layered superconductors.


\vspace{5mm}

\begin{acknowledgments}
$Acknowledgments$: This work made use of the Cornell Center for Materials Research shared instrumentation facility. H.Y. and D.A.M. acknowledge support by the NSF Platform for the Accelerated Realization, Analysis, and Discovery of Interface Materials (PARADIM) under cooperative agreement No. DMR-2039380. The materials synthesis part of this work was supported by the U.S. Department of Energy, Office of Science, Office of Basic Energy Sciences under Award Number DE-SC0026002 and by the Gordon and Betty Moore Foundation’s EPiQS Initiative (grant numbers GBMF3850 and GBMF9073).
\end{acknowledgments}

\bibliography{MEP_O_dopant}
\end{document}